\documentclass[%
 reprint,showkeys,
 amsmath,amssymb,
 aps,
]{revtex4-2}

\usepackage{graphicx}
\usepackage{dcolumn}
\usepackage[usenames]{color}
\usepackage{colortbl}
\usepackage{bm}
\usepackage{longtable}

\begin{document}

\title{Photoproduction of the ${^{55-57}}$Co nuclei on ${^{\rm nat}}$Ni \\
	at the bremsstrahlung end-point energy of 35--94 MeV }


\author{I.S. Timchenko$^{1,2,}$} \email{iryna.timchenko@savba.sk;\\ timchenko@kipt.kharkov.ua}
\author{O.S. Deiev$^2$, S.M. Olejnik$^2$, S.M. Potin$^2$, \\
	V.A. Kushnir$^2$, V.V. Mytrochenko$^{2}$, S.A. Perezhogin$^2$, A. Herz\'{a}\v{n}$^1$}

\affiliation{$^1$ Institute of Physics, Slovak Academy of Sciences, SK-84511 Bratislava, Slovakia}%
\affiliation{$^2$ National Science Center "Kharkov Institute of Physics and Technology", \\ 1, Akademichna St., 61108, Kharkiv, Ukraine}%

\date{\today}

\begin{abstract}
 	Production of the ${^{55-57}}$Co nuclei on ${^{\rm nat}}$Ni in photonuclear reactions using bremsstrahlung gamma photon irradiation with end-point energy $E_{\rm{\gamma max}}$ between 35 and 94~MeV has been studied. The experiment was performed at the electron linear accelerator LUE-40 NSC KIPT using the  methods of $\gamma$ activation and off-line $\gamma$-ray spectroscopy. The obtained experimental flux-averaged cross-sections $\langle{\sigma(E_{\rm{\gamma max}})}\rangle$ agree with the data found in  literature. 
The theoretical flux-averaged cross-sections $\langle{\sigma(E_{\rm{\gamma max}})}\rangle_{\rm{th}}$ for the production of ${^{55-57}}$Co and ${^{55-57}}$Ni were estimated using the cross-section values $\sigma(E)$ from the TALYS1.95 code and bremsstrahlung spectra of gamma photons calculated by GEANT4.9.2.
The experimental results for ${^{56,57}}$Co agree with the cumulative $\langle{\sigma(E_{\rm{\gamma max}})}\rangle_{\rm{th}}$. 
However theoretical prediction fails to reproduce the measured cross-sections for the production of ${^{55}}$Co.
	\end{abstract}
\maketitle

\section{Introduction}

Most of the data on cross-sections for photonuclear reactions are important for traditional studies of the Giant Dipole Resonance (GDR), mechanisms of its excitation and decay including competition between statistical and direct processes in decay channels, GDR configurational and isospin splitting, sum rule exhaustion, etc. Experimental cross-sections for photonuclear reactions are also widely used in various applications, primarily in astrophysics \cite{astro}, medicine \cite{med}, design of fast reactors \cite{FR} and accelerator driven sub-critical systems \cite{ADS1,ADS2}. 

Such data were obtained in experiments using high-intensity $\gamma$-fluxes either quasi-monoenergetic annihilation photons \cite{1,2,3,4,5,6}, or high-energy electron bremsstrahlung \cite{7,8,9,10}. Data on these cross-sections can be found in the international digital databases \cite{12}. 

Despite the significant amount of experimental data available in international databases, there is still an insufficiency of data for the cross-sections of many-particle photoneutron reactions. Mainly, data for reactions resulting in the release of up to 4 neutrons can be found. In addition, experimental data from different laboratories do not always agree with each other, even in the GDR range \cite{a11,a12,16}. Data for photonuclear reactions involving charged particles in the output channel are mainly limited to the region of light nuclei and energies within the GDR range.

Experimental research into photonuclear reactions is currently being performed in several directions. Thus, in a number of works, previously measured results are revised and criteria of data reliability are developed \cite{15,16}. Repeated systematic measurements of cross-sections for previously studied reactions are performed, as noted in \cite{17}. Studies are being carried out on high-threshold photoneutron reactions \cite{18,19}, photoproton \cite{Hf180,Mo_95Nb}, and photonuclear reactions with the emission of charged particles/clusters \cite{20,21,22}.

Theoretical estimates of cross-sections for photonuclear reactions are commonly performed now using the TALYS code \cite{Ko2008,Ko2017}. This code is constantly being improved and is used to simulate the processes of interaction of particles with the atomic nuclei. Additionally, it is employed in calculations relating to astrophysics, fission yields, medical isotopes production, and thermal scattering data. The results for the output files from TALYS, which are based on default parameters, adjusted calculations and data from different sources, are placed in the TENDL tables \cite{TENDL}.

It should be noted that theoretical estimates of photonuclear cross-sections in TALYS do not always agree with experimental data. Discrepancies occur for many-particle photoneutron reactions \cite{18,19}, as well as in the case of reactions with the emission of charged particles (including clusters) \cite{Hf180,Mo_95Nb,21,22}.

The photonuclear cross-sections on $^{\rm nat}$Ni are important because nickel is of practical interest as an important structural and surface material, which is often used in structural materials for accelerator and nuclear technology \cite{a1}. The physical and mechanical properties of nickel are suitable for the manufacture of heat-resistant and corrosion-resistant alloys \cite{a2}. Nickel is an important structural material not only for modern reactors, but also for generation IV reactors and future fusion reactors \cite{a3}. From a medical and research point of view, the photon-induced reaction cross-sections of Co and Ni can be used to produce important isotopes such as ${^{55-58}}$Co, ${^{56}}$Ni, and ${^{57}}$Ni due to their suitable decay characteristics. For example, the ${^{55}}$Co isotope is useful for diagnostics by positron emission tomography, and could be used for the diagnostics of lung cancer and for the visualization of tumours \cite{a4}. The ${^{57}}$Co isotope 
	general use is as a calibration standard for $\gamma$-spectrometers and single photon emission
	tomographs \cite{a4}, the ${^{56}}$Co isotope is used as a calibration source up to 3.5~MeV for $\gamma$-ray detectors \cite{a5}. 

Experimental data on cross-sections for photonuclear reactions on nickel at energies of 12--40 MeV were obtained in a number of studies \cite{23,24,25,26,27,28,29,30,31,32,33,Zhe,Ishch}. In \cite{15}  estimated cross-section values for the reactions of photoproduction of Co nuclei on ${^{58,60}}$Ni are presented, and obtained on the basis of the model of the phenomenological description of the main GDR decay channel competition at energies of 10--40 MeV. At higher energies, the study of cross-sections on natural nickel using electron bremsstrahlung was performed in \cite{34,35}. The flux-averaged cross-sections of $^{\rm nat}$Ni($\gamma$,$x$np)$^{55-58}$Co reactions were measured in these studies using the $\gamma$-activation technique at bremsstrahlung end-point energies of 55--65~MeV \cite{34} and 65--75~MeV \cite{35}.  

This paper presents results of the experimental study of the $^{55-57}$Co production in photonuclear reactions on natural nickel measured using bremsstrahlung spectra of gamma photons with end-point energy $E_{\rm{\gamma max}}$ between 35 and 94 MeV.
The obtained data are compared both with the theoretical calculation performed using partial cross-sections $\sigma(E)$ from the TALYS1.95 code \cite{Ko2008,Ko2017}, and with the experimental data available in the literature \cite{34,35,Zhe}.

\section{Experimental setup and procedure}
\label{sec:1}

\begin{figure}[ht!]
	\resizebox{0.52\textwidth}{!}{%
		\includegraphics{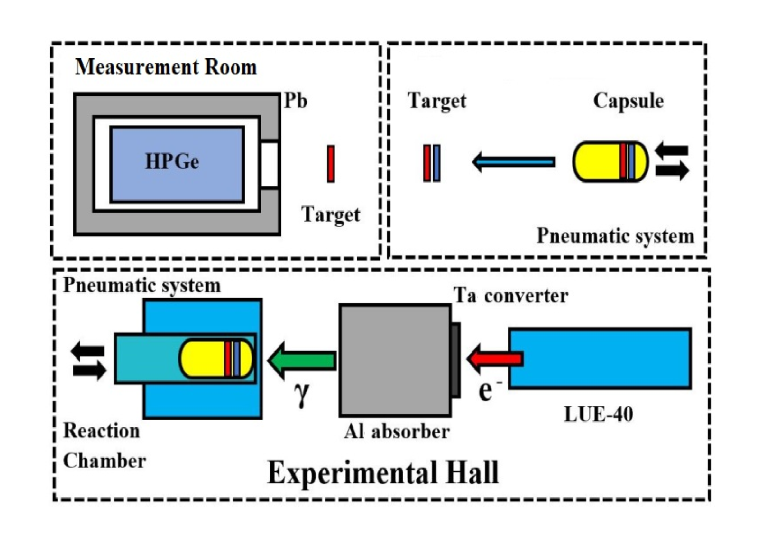}}
	\caption{Schematic block diagram of the experiment. The upper part shows the measurement room. The lower part shows the electron accelerator LUE-40, Ta converter, Al absorber, and exposure reaction chamber. }
	\label{fig1}
\end{figure}

The photoproduction of the  $^{55-57}$Co nuclei was experimentally studied via measurement of residual $\gamma$-activity of the irradiated targets. This technique has been described in many studies of photonuclear reactions, e.g. \cite{44,45,46,47}. The schematic block diagram of the experiment is presented in Fig.~\ref{fig1}. 

In the experiment, the LUE-40 electron linear accelerator of Research and Development Center "Accelerator" of the National Science Center "Kharkov Institute of Physics and Technology" \cite{48, 49} was employed. To generate the bremsstrahlung $\gamma$-flux, electrons with the initial energy $E_{\rm e}$ were impinging  on the converter made of a natural tantalum square plate of 20~mm and a thickness of 1.05 mm. To  stop  the electrons,  a cylindrical aluminum absorber, 100 mm in diameter and 150 mm in length, was used. The studied target and target-monitor, placed in the thin aluminum capsule, were positioned behind the Al absorber on the electron beam axis. For transporting the targets to the irradiation point and back for induced activity measurement, the pneumatic tube transfer system was used. After the irradiated targets are delivered to the measurement room, the targets are extracted from the aluminum capsule and transferred one by one to the detector for the measurements. The semiconductor High-Purity Germanium (HPGe) detector with the energy resolution (FWHM) of 0.8 and 1.8~keV at 122 and 1332.5~keV, respectively, was used. Its detection efficiency, $\varepsilon$, at 1332.5~keV was 20\% relative to the NaI(Tl) scintillator, 3~inches in diameter and 3~inches in thickness. The detector was equipped with passive Pb shielding to suppress natural background. The  detection efficiency calibration was done using a standard set of $\gamma$ sources: $^{22}$Na, $^{60}$Co, $^{133}$Ba, $^{137}$Cs, $^{152}$Eu, and $^{241}$Am.

To investigate the photonuclear reactions of the production of $^{55-57}$Co nuclei on nickel, targets made of natural Ni were used. Isotopic composition of the $^{\rm nat}$Ni material (\%) is: $^{58}$Ni -- 68.077, $^{60}$Ni -- 26.223, $^{61}$Ni -- 1.140, $^{62}$Ni -- 3.634, $^{64}$Ni -- 0.926 \cite{Ko2017,50}.  The admixture of other elements in the targets did not exceed 0.1\% by weight. The targets had a disk-like shape with a diameter of 8 mm and a weight of $\sim$80 mg. 

The bremsstrahlung $\gamma$-flux was monitored by the yield of the $^{100}$Mo$(\gamma,{\rm n})^{99}$Mo reaction and calculated using the GEANT4.9.2 code \cite{50,51,52}.
For this purpose, the $^{\rm nat}$Mo and the $^{\rm nat}$Ni targets were closely placed to each other and simultaneously irradiated. The $^{\rm nat}$Mo targets had a form of 8 mm diameter disks with a mass of approximately 60~mg. The monitoring procedure was performed by comparing the measured flux-averaged cross-section values with the calculated data using the procedure described in \cite{10}. In the case of the $^{100}$Mo isotope, the isotopic abundance of  9.63~\% \cite{Ko2017,50} was used in our calculations.  

In the experiment, the $^{\rm nat}$Ni and $^{\rm nat}$Mo targets were exposed to bremsstrahlung radiation with  end-point energy $E_{\rm{\gamma max}}$ ranging from 35 to 94~MeV with an energy step of  $\sim$5~MeV. The irradiation time $t_{\rm irr}$ was 30 min for each  $E_{\rm{\gamma max}}$ value. 

During irradiation of the $^{\rm nat}$Ni targets, products of the ($\gamma$,$x$n$y$p) reaction channels and their decay products accumulate in the targets. The $^{55-57}$Co nuclei are formed  as an result of the $^{\rm nat}$Ni($\gamma$,$x$np) reaction. Also after the EC/$\beta^+$~(100\%) decay of the $^{55-57}$Ni nuclei, which are formed in targets as a result of ($\gamma$,$x$n)  reactions. Accordingly, in the case of the study of the $^{55-57}$Co nuclei, we need to take into account two production channels of the accumulation. 
	Decay characteristics of the $^{55-57}$Co and $^{55-57}$Ni nuclei are listed in Table~\ref{tab1} according to \cite{53}. In addition, also data for the monitor reaction $^{100}$Mo$(\gamma,{\rm n})^{99}$Mo are given.

Measurement time, $t_{\rm meas}$, of the residual $\gamma$-activity of the $^{\rm nat}$Mo targets was 30 min.
Before measuring the $^{55}$Co activity, the $^{\rm nat}$Ni targets were first cooled for a minimum of 60~min, and then measured for 30 min. The prolonged cooling time of 32 up to 53 days was chosen in the case of the $^{56,57}$Co photoproduction study, and the energy spectra of the induced activity were measured for 20 up to 95 hours.

The uncertainty in measurements of experimental values of the flux-averaged cross-sections $\langle{\sigma(E_{\rm{\gamma max}})}\rangle$ was determined as a quadratic sum of statistical and systematical errors. 

The systematical errors are due to the following uncertainties: 1) exposure time and the electron current $\sim$0.5\%; 2) detection efficiency of the HPGe detector $\sim$2--3\%, which is generally associated with the radiation source error and the choice of the approximation curve; 3) the half-life, $T_{1/2}$, of the reaction products and the intensity, $I_{\gamma}$, of the analyzed transition are noted in Table~\ref{tab1}; 4) the number of target atoms $\sim$0.2\%; 5) normalization of the experimental data to the yield of the monitoring reaction $^{100}$Mo$(\gamma,{\rm n})^{99}$Mo made up to 6\%. 

The statistical error in the observed $\gamma$-activity is mainly due to statistics calculation in the full absorption peaks of the corresponding transitions, which varies from 0.15\% to 7.65\%. The peak area measurements showed the highest statistics in the production of the $^{57}$Co nucleus, with a statistical error of 0.15--0.65\%. The most difficult measurements were performed for the $^{55}$Co nucleus, with a statistical error of 2.15--7.65\%.

  \begin{table*}[]
		\caption{\label{tab1} Spectroscopic data of the products of specific reaction channels adopted from \cite{53}: spin ($J$) and parity ($\pi$) of the ground state of the studied Nucleus, their half-life ($T_{1/2}$); Decay mode is the type of decay of the studied Nucleus, $E_{\gamma}$ and $I_{\gamma}$ are the energies and intensities of the $\gamma$-ray transitions associated with the decay of the nucleus, respectively. In the column "Contributing reactions" lists of possible reactions of production $^{55-57}$Co and $^{55-57}$Ni on stable isotopes of $^{\rm nat}$Ni. $E_{\rm{thr}}$ denotes threshold energy of the reactions (calculated by TALYS1.95 \cite{Ko2008,Ko2017}). 
		}
		\centering
		\begin{tabular}{ccccccc}		\hline
			\begin{tabular}{c} Nucleus  ($J^{\pi}$) \end{tabular} & $T_{1/2}$  & \begin{tabular}{c} Decay mode (\%)
			\end{tabular} & $E_{\gamma}$ (keV)  & $I_{\gamma}$ (\%) & \begin{tabular}{c} Contributing \\ reactions
			\end{tabular} &  $E_{\rm{thr}}$ (MeV) \\ 	\hline
			&&&&&&\\ 	   
			$^{55}$Co ($7/2^{-}$) & 17.53~{\it  3} h & EC/$\beta^+$ (100\%) & 931.1  & 75~{\it  4}  & $^{58}\rm{Ni}(\gamma,2np)$ & \textbf{29.63} \\
			&&&&&  $^{60}\rm{Ni}(\gamma,4np)$ & 50.02 \\
			&&&&&  $^{61}\rm{Ni}(\gamma,5np)$ & 57.84 \\
			&&&&&  $^{62}\rm{Ni}(\gamma,6np)$ & 68.43 \\
			&&&&&&\\ 
			$^{56}$Co ($4^{+}$) & 77.236~{\it  26} d & EC/$\beta^+$ (100\%) & 846.77  & 99.9399~{\it  23} & $^{58}\rm{Ni}(\gamma,np)$ & \textbf{19.55} \\
			&&& 1037.84  & 14.05~{\it  4}  &  $^{60}\rm{Ni}(\gamma,3pn)$ & 39.94 \\
			&&& 1238.28  & 66.46~{\it  12}  &  $^{61}\rm{Ni}(\gamma,4np)$ & 47.76 \\
			&&&&&  $^{62}\rm{Ni}(\gamma,5np)$ & 58.35 \\
			&&&&&  $^{64}\rm{Ni}(\gamma,7np)$ & 74.85 \\
			&&&&&&\\ 
			$^{57}$Co ($7/2^{-}$) & 271.74~{\it  6} d & EC (100\%) & 122.06  & 85.60~{\it  17} & $^{58}\rm{Ni}(\gamma,np)$ & \textbf{8.17} \\
			&&& 136.47  & 10.68~{\it  8}  &  $^{60}\rm{Ni}(\gamma,2np)$ & 28.56 \\
			&&&   &   &  $^{61}\rm{Ni}(\gamma,3np)$ & 36.38 \\
			&&&&&  $^{62}\rm{Ni}(\gamma,4np)$ & 46.98 \\
			&&&&&  $^{64}\rm{Ni}(\gamma,6np)$ & 63.47 \\
			&&&&&&\\ 
			$^{55}$Ni ($7/2^{-}$) & 204.7~{\it  37} ms & EC/$\beta^+$ (100\%) & --  & -- & $^{58}\rm{Ni}(\gamma,3n)$ & \textbf{39.11} \\
			&&&    &  &  $^{60}\rm{Ni}(\gamma,5n)$ & 59.49 \\
			&&&   &   &  $^{61}\rm{Ni}(\gamma,6n)$ & 67.31 \\
			&&&&&&\\ 
			$^{56}$Ni ($0^{+}$) & 6.075~{\it  10} d & EC/$\beta^+$ (100\%) & 158.38  & 98.8~{\it  10} & $^{58}\rm{Ni}(\gamma,2n)$ &\textbf{22.46} \\
			&&&    &  &  $^{60}\rm{Ni}(\gamma,4n)$ & 42.85 \\
			&&&   &   &  $^{61}\rm{Ni}(\gamma,5n)$ & 50.67 \\
			&&&   &   &  $^{62}\rm{Ni}(\gamma,6n)$ & 61.27 \\
			&&&&&&\\ 
			$^{57}$Ni ($3/2^{-}$) & 35.60~{\it  6} h & EC/$\beta^+$ (100\%) & 127.16  & 16.7~{\it  5} & $^{58}\rm{Ni}(\gamma,n)$ & \textbf{12.22} \\
			&&& 1377.63  & 81.7~{\it  24} &  $^{60}\rm{Ni}(\gamma,3n)$ & 32.60 \\
			&&&   &   &  $^{61}\rm{Ni}(\gamma,4n)$ & 40.42 \\
			&&&   &   &  $^{62}\rm{Ni}(\gamma,5n)$ & 51.02 \\
			&&&   &   &  $^{64}\rm{Ni}(\gamma,6n)$ & 67.51 \\
			&&&&&&\\ 
			$^{99}$Mo ($1/2^{+}$) &  65.924~{\it  6} h & $ \beta^-$ (100\%) & 739.50 & 12.2~{\it  2}  & $^{100}\rm{Mo}(\gamma,n)$ & \textbf{8.29} \\ \hline
		\end{tabular}
	\vspace{-3ex}
	\end{table*}

	\section{  Calculation of flux-averaged cross-sections  }
	\label{sec:2}
	
	The cross-section  $\langle{\sigma(E_{\rm{\gamma max}})}\rangle$  averaged over the bremsstrahlung $\gamma$-flux $W(E,E_{\rm{\gamma max}})$ from the threshold energy, $E_{\rm thr}$, of the studied reaction to the end-point energy of the spectrum, $E_{\rm{\gamma max}}$, was calculated by equation: 
	
	\begin{equation}\label{form1}
		\langle{\sigma(E_{\rm{\gamma max}})}\rangle = 
		\frac{ \int\limits_{E_{\rm{thr}}}^{E_{\rm{\gamma max}}}\sigma(E) \cdot  W(E,E_{\rm{\gamma max}})dE}{{\int\limits_{E_{\rm{thr}}}^{E_{\rm{\gamma max}}} W(E,E_{\rm{\gamma max}})dE}}.
	\end{equation}
	
	To calculate the flux-averaged cross-section of the production of a particular nucleus on a natural element that has several stable isotopes, Eq.~(\ref{form1}) must be transformed into a sum of the flux-averaged cross-sections  $\langle{\sigma(E_{\rm{\gamma max}})}\rangle_i$ of the production of the nucleus on each of the isotopes $i$, where each has its own specific reaction threshold $E^i_{\rm thr}$:
	
	\begin{equation*}\label{form1a}
		\langle{\sigma(E_{\rm{\gamma max}})}\rangle = \sum^m_{i=1}
		\frac{ \int\limits_{E^i_{\rm{thr}}}^{E_{\rm{\gamma max}}} A_i \cdot  \sigma_i(E) \cdot  W(E,E_{\rm{\gamma max}})dE}{{\int\limits_{E^i_{\rm{thr}}}^{E_{\rm{\gamma max}}} W(E,E_{\rm{\gamma max}})dE}},     \qquad\qquad\qquad\qquad (1\rm a)
	\end{equation*}
	
	where $m$ is the number of stable isotopes of natural element, on which a 
	nucleus of interest can be formed. In the case of $^{\rm nat}$Ni and the production of $^{55-57}$Co nuclei, $m = 5$. $A_i$ is the abundance of the $i$-th stable isotope.
	
	The flux-averaged cross-section $\langle{\sigma(E_{\rm{\gamma max}})}\rangle$  can also be determined as the sum of the flux-averaged cross-sections on all stable isotopes $i$-$m$, which are weighted with the integral flux for the minimum reaction threshold $E^{\rm min}_{\rm thr}$:
	
	\begin{equation*}\label{form1b}
		\langle{\sigma(E_{\rm{\gamma max}})}\rangle = 
		\frac{ \sum^m_{i=1} \int\limits_{E^i_{\rm{thr}}}^{E_{\rm{\gamma max}}} A_i \cdot  \sigma_i(E) \cdot  W(E,E_{\rm{\gamma max}})dE}{{\int\limits_{E^{\rm min}_{\rm{thr}}}^{E_{\rm{\gamma max}}} W(E,E_{\rm{\gamma max}})dE}}.\qquad\qquad\qquad\qquad (1\rm b)
	\end{equation*}
	
	The difference between threshold energies $E^i_{\rm{thr}}$ and $E^{\rm min}_{\rm{thr}}$ leads to different magnitudes ofthe effective integral bremsstrahlung $\gamma$-flux and  $\gamma$-flux from $E^{\rm min}_{\rm thr}$, respectively. As a result, the estimates given by Eq.~(1a) and Eq.~(1b)  differ from each other. 
	
	Experimental $\langle{\sigma(E_{\rm{\gamma max}})}\rangle$ cross-sections are calculated as following:
	
	\begin{equation}
		\langle{\sigma(E_{\rm{\gamma max}})}\rangle = 
		\frac{\lambda \triangle A  {\rm{\Phi}}^{-1}(E_{\rm{\gamma max}})}{N_n I_{\gamma} \ \varepsilon (1-e^{-\lambda t_{\rm{irr}}})e^{-\lambda t_{\rm{cool}}}(1-e^{-\lambda t_{\rm{meas}}})},
		\label{form2}
	\end{equation}
	
	where $\triangle A$ is the number of counts in the full absorption peak, $\lambda$ is the decay constant (s$^{-1}$),  $N_n$ is the number of atoms of the natural nickel target, $I_{\gamma}$ is the intensity of the measured $\gamma$ quanta, $\varepsilon$ is the absolute detection efficiency at the measured $\gamma$ quanta energy, ${\rm{\Phi}}(E_{\rm{\gamma max}}) = {\int\limits_{E^{\rm  min}_{\rm{thr}}}^{E_{\rm{\gamma max}}}W(E,E_{\rm{\gamma max}})dE}$ is the integrated flux of bremsstrahlung gamma photons in the energy range from the minimum reaction threshold $E^{\rm min}_{\rm thr}$ up to $E_{\rm{\gamma max}}$; $t_{\rm{irr}}$, $t_{\rm{cool}}$ and $t_{\rm{meas}}$ are the irradiation time, cooling time and measurement time, respectively. A more detailed description of all the calculation procedures necessary for the determination of  $\langle{\sigma(E_{\rm{\gamma max}})}\rangle$ can be found in \cite{10,19}. 
	
	The reaction yield is defined as:
	
	\begin{equation}\label{form3}
		Y(E_{\rm{\gamma max}}) = N_n \int\limits_{E_{\rm{thr}}}^{E_{\rm{\gamma max}}}\sigma(E)\cdot W(E,E_{\rm{\gamma max}})dE.
	\end{equation}
	
	To estimate a reaction contribution to the total production of a particular nuclide (e.q., the $^{58}$Ni($\gamma$,p) reaction in the production of the $^{57}$Co nucleus on $^{\rm nat}$Ni), the normalized reaction yield $Y_k(E_{\rm{\gamma max}})$ was used. The $Y_k(E_{\rm{\gamma max}})$ yield  can be calculated as follows: 
	
	\begin{equation}\label{form4}
		Y_k(E_{\rm{\gamma max}}) = \frac
		{A_k \int\limits_{E^k_{\rm{thr}}}^{E_{\rm{\gamma max}}}\sigma_k(E)\cdot W(E,E_{\rm{\gamma max}})dE}
		{\sum^m_{i=1} A_i \int\limits_{E^i_{\rm{thr}}}^{E_{\rm{\gamma max}}}\sigma_i(E)\cdot W(E,E_{\rm{\gamma max}})dE}.
	\end{equation}
	
	Equation~(\ref{form4}) is useful when the flux-averaged cross-section of the dominant reaction for the production of the studied nucleus on one of the isotopes in the natural element need to be obtained. For this purpose, it is necessary to correct the values calculated by Eq.~(\ref{form2}) by using the corresponding value of the normalized reaction yield $Y_k(E_{\rm{\gamma max}})$ and the isotopic abundance in the target material.

	\section{Results and discussion}
	\label{sec:4}
	\subsection{Calculated $\sigma(E)$ and flux-averaged $\langle{\sigma(E_{\rm{\gamma max}})}\rangle$ cross-sections }
	\label{sec:4a}
	
	The  experimental cross-sections of photonuclear production of the $^{55-57}$Co and $^{55-57}$Ni nuclei on stable nickel isotopes, existing early and obtained in this study, are compared with theoretical predictions. In this work, the calculation of the theoretical cross-sections $\sigma(E)$  was performed using the TALYS1.95 code \cite{Ko2008,Ko2017} with default options.
	
\begin{figure*}[ht!]
		\begin{minipage}[h]{0.45\linewidth}
			\center{\includegraphics[width=1\linewidth]{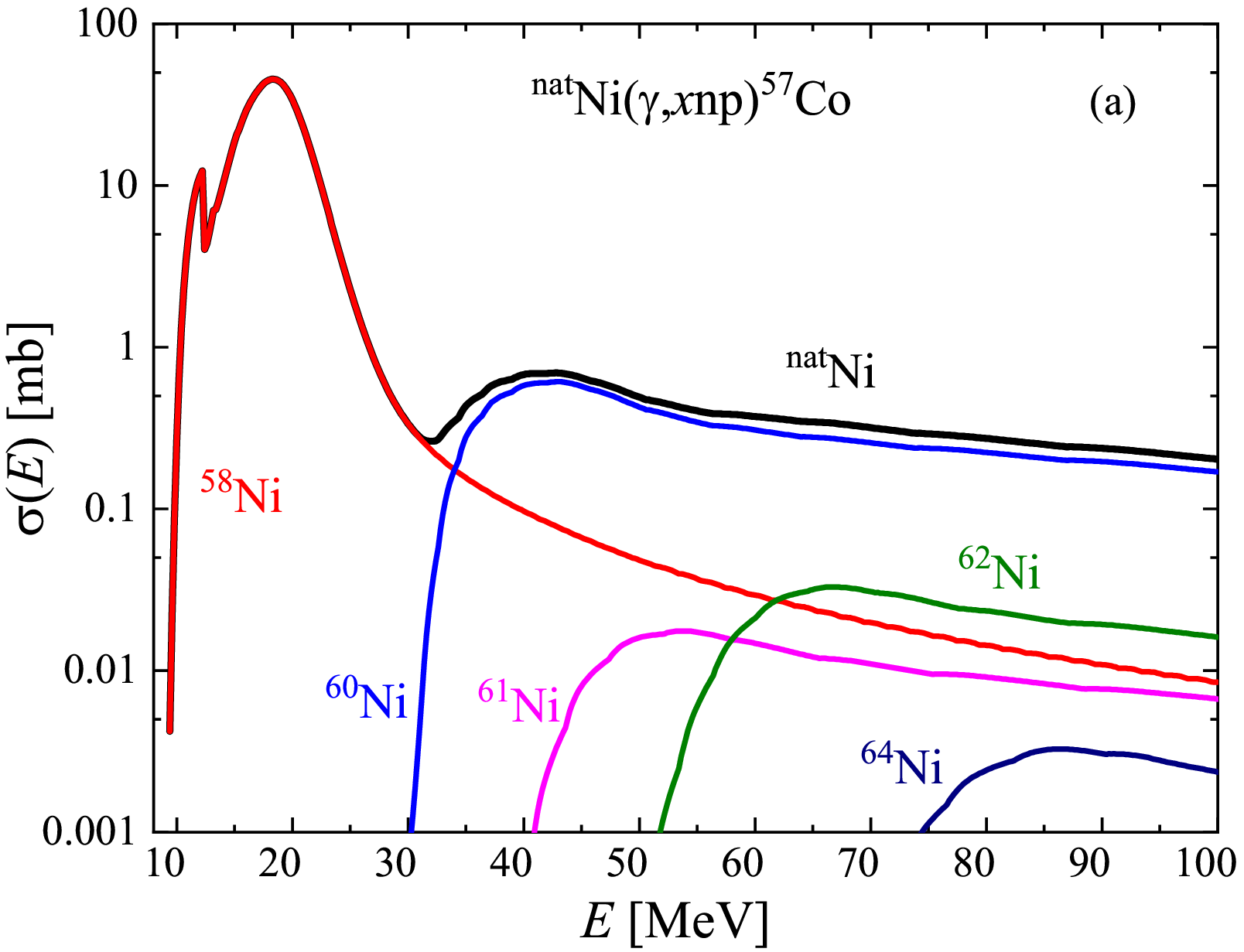}} \\
		\end{minipage}
		\hfill
		\begin{minipage}[h]{0.45\linewidth}
			\center{\includegraphics[width=1\linewidth]{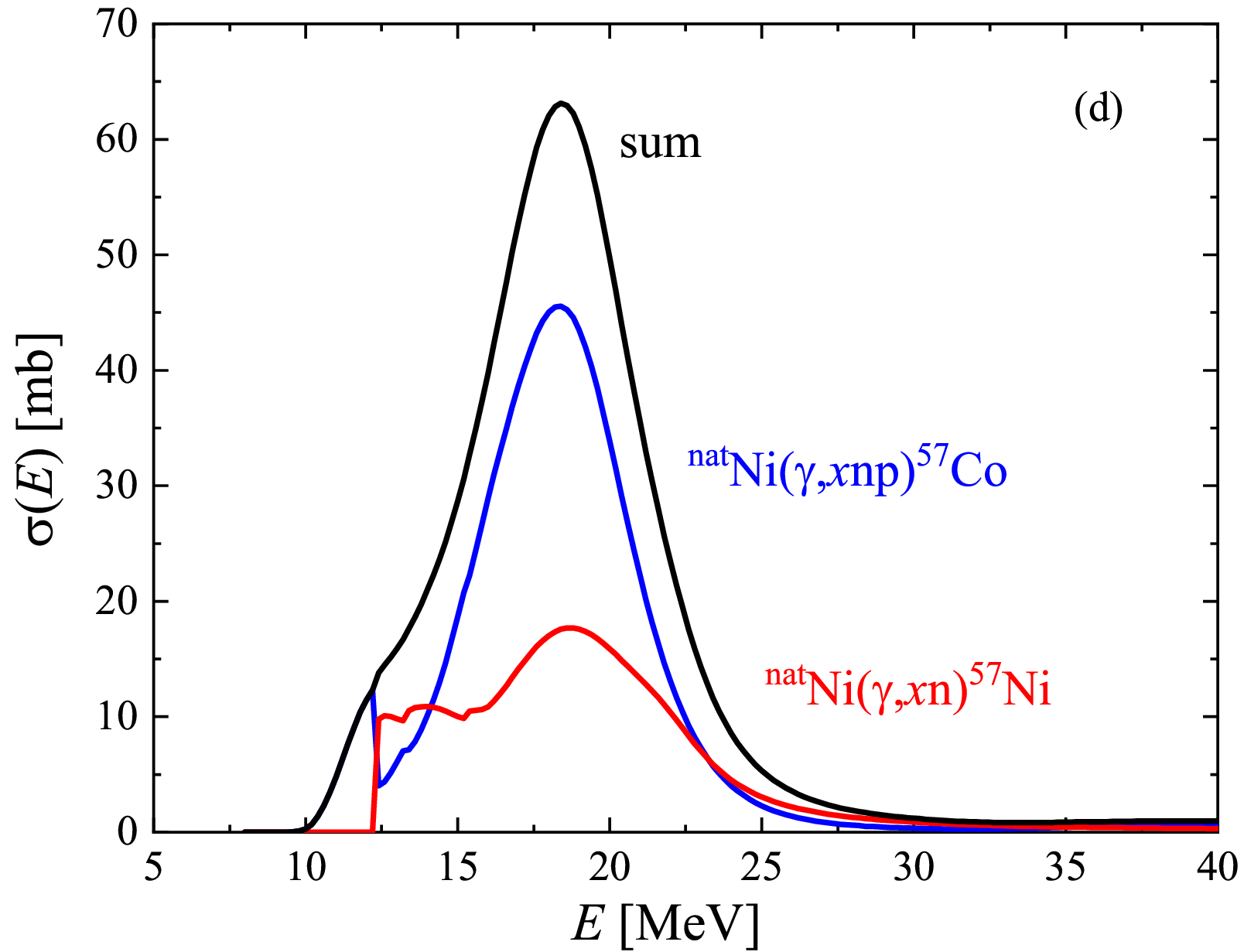}} \\
		\end{minipage}
		\vfill
		\begin{minipage}[h]{0.45\linewidth}
			\center{\includegraphics[width=1\linewidth]{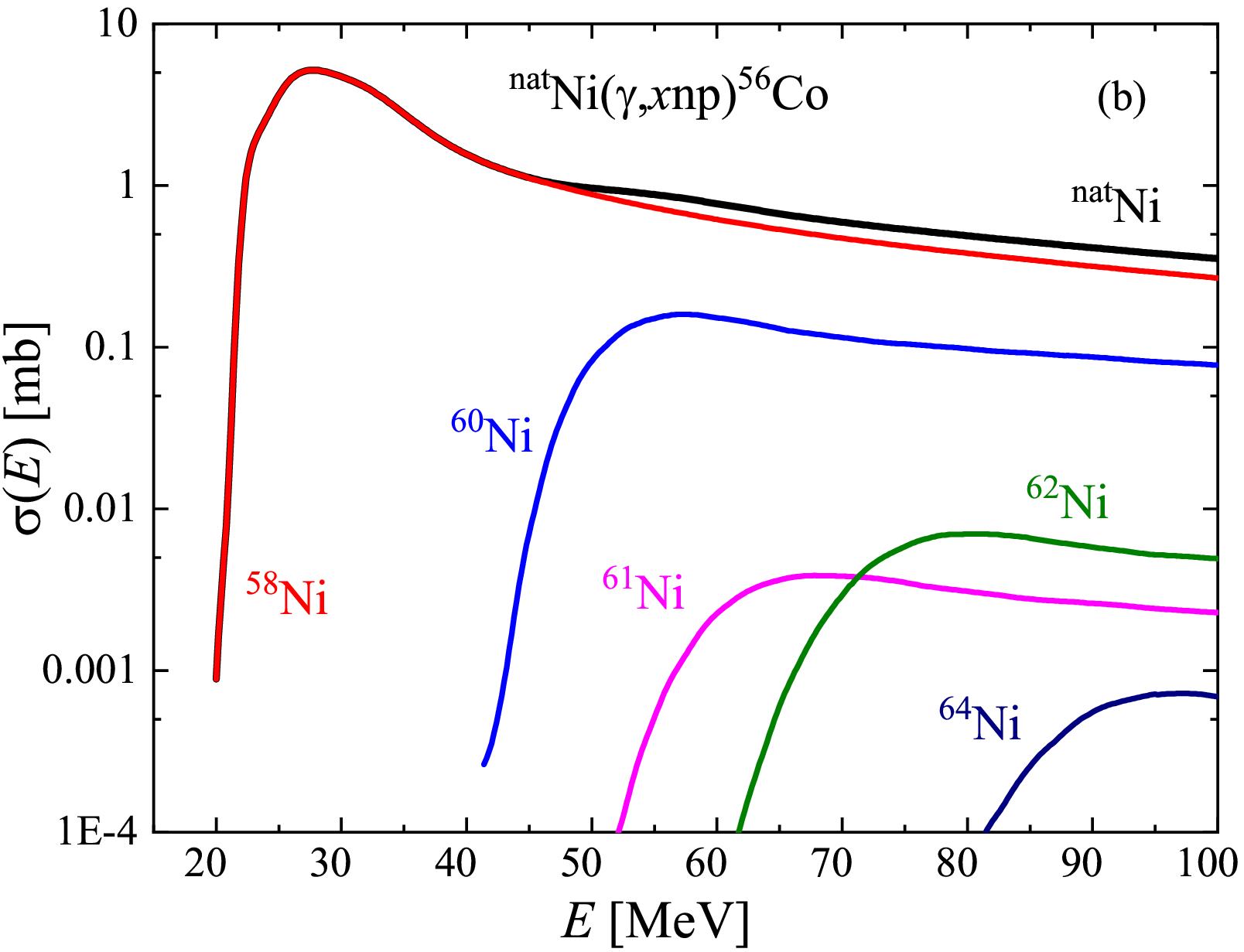}} \\
		\end{minipage}
		\hfill
		\begin{minipage}[h]{0.45\linewidth}
			\center{\includegraphics[width=1\linewidth]{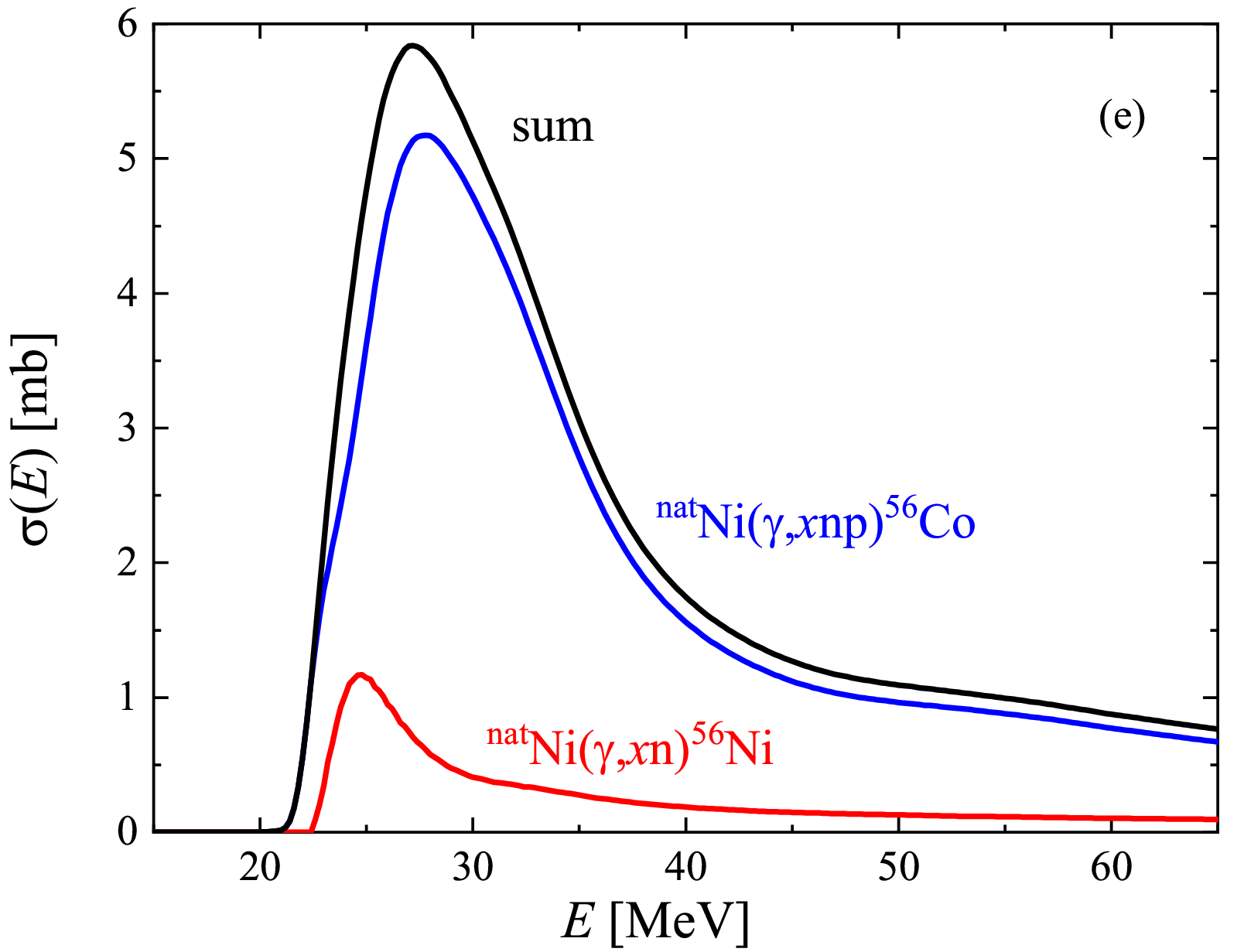}} \\
		\end{minipage}
		\begin{minipage}[h]{0.45\linewidth}
			\center{\includegraphics[width=1\linewidth]{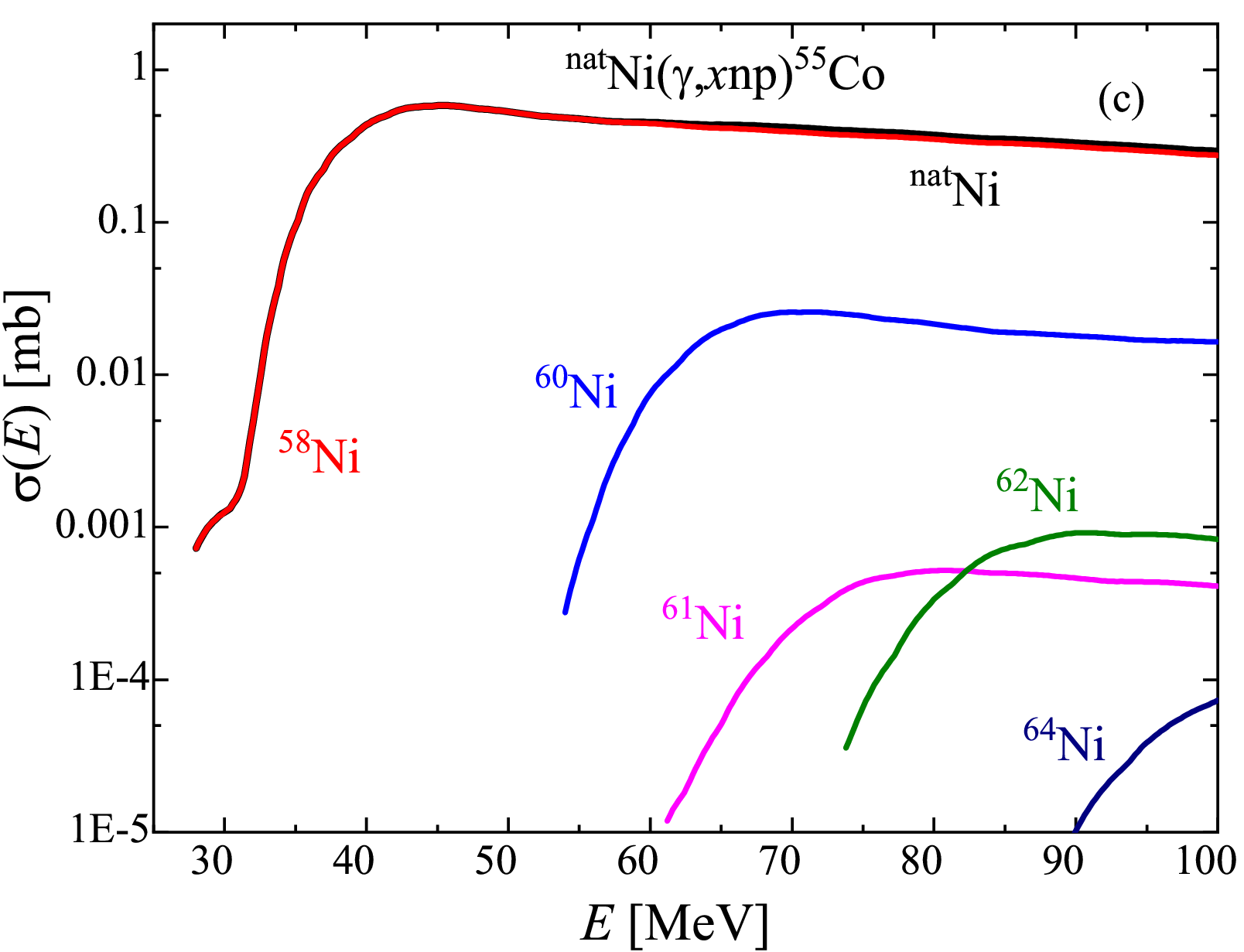}} \\
		\end{minipage}
		\hfill
		\begin{minipage}[h]{0.45\linewidth}
			\center{\includegraphics[width=1\linewidth]{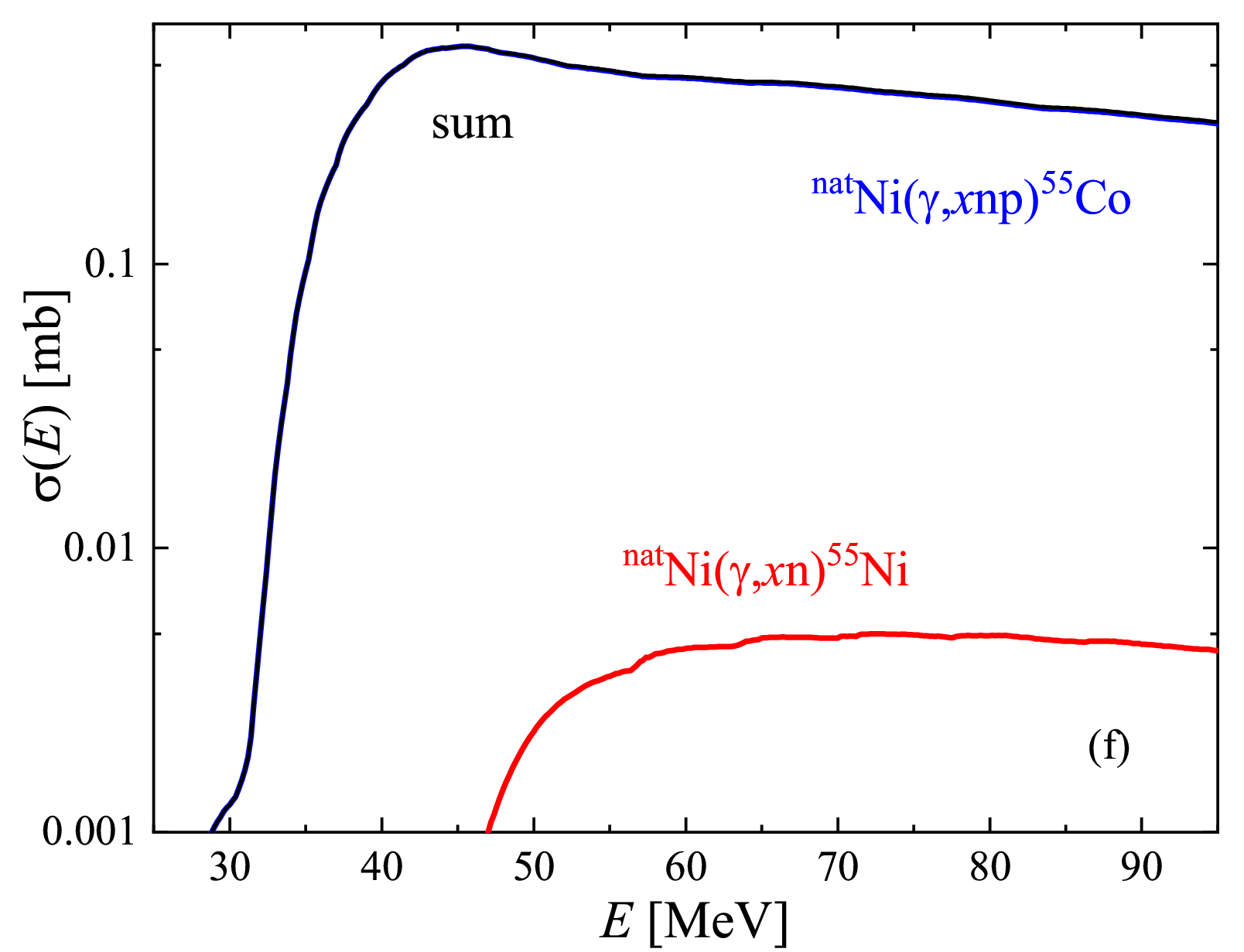}} \\
		\end{minipage}
		\caption{The theoretical cross-sections $\sigma(E)$ from the TALYS1.95 code.\\
			(a--c) The cross-sections of the photoproduction of $^{55-57}$Co nuclei on stable nickel isotopes, taking into account the isotopic abundance of the elements. The production of the studied nucleus on $^{\rm nat}$Ni is a sum of the cross-sections for each Ni isotope.\\
			(d--f) The total cross-sections $\sigma(E)$ for the photoproduction of $^{55-57}$Co and $^{55-57}$Ni nuclei on natural nickel. The black curves are the sums of the $^{\rm nat}$Ni($\gamma$,$x$np)$^{55-57}$Co and $^{\rm nat}$Ni($\gamma$,$x$n)$^{55-57}$Ni cross-sections. }
		\label{fig2}
	\end{figure*}

	Figure~\ref{fig2} shows the calculated cross-sections $\sigma(E)$ for the production of $^{57}$Co (a), $^{56}$Co (b) and $^{55}$Co (c) nuclei on stable nickel isotopes, taking into account the isotopic abundance of the elements. To calculate the cross-section for the production of the studied nucleus on $^{\rm nat}$Ni, we summed the cross-sections for each Ni isotope.  
	
	The total cross-sections $\sigma(E)$ for the photoproduction of $^{57}$Co and $^{57}$Ni, $^{56}$Co and $^{56}$Ni, $^{55}$Co and $^{55}$Ni nuclei on natural nickel, respectively, are shown in Figs.~\ref{fig2} (d-f). The sum of cross-sections of the $^{\rm nat}$Ni($\gamma,x$n)$^{55-57}$Ni and $^{\rm nat}$Ni($\gamma$,$x$np)$^{55-57}$Co reactions represents the cumulative yield of the $^{55-57}$Co nuclei production on $^{\rm nat}$Ni. The contribution of the $^{\rm nat}$Ni($\gamma,x$n)$^{55-57}$Ni photonuclear reactions is maximal in the case of the $^{57}$Co production and negligible in the case of $^{55}$Co.
	
	\subsection{The experimental $\langle{\sigma(E_{\rm{\gamma max}})}\rangle$ cross-sections of the $^{55-57}$Co nuclei production on $^{\rm nat}$Ni }
	\label{sec:4b}
	
\begin{figure*}[ht!]
		\begin{minipage}[h]{1\linewidth}
			{\includegraphics[width=1\linewidth]{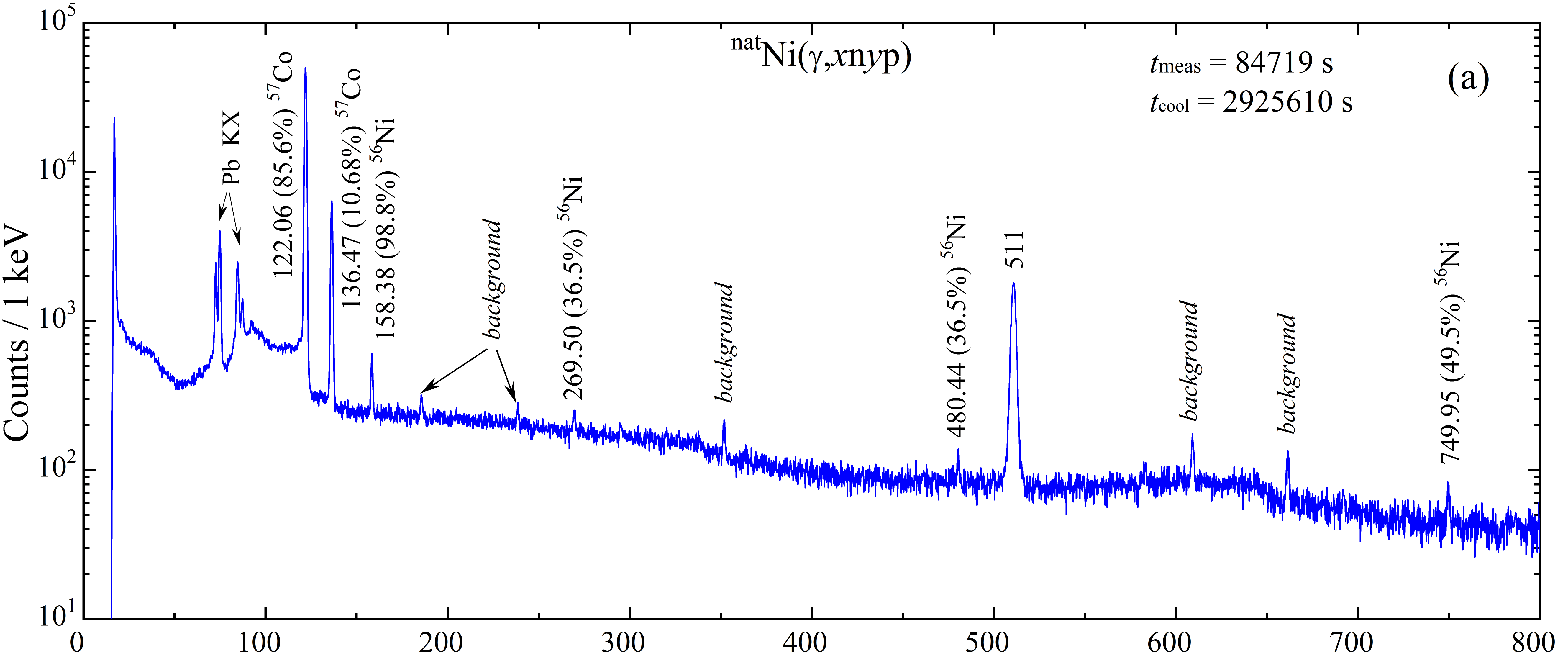}} \\
		\end{minipage}
		\vfill
		\begin{minipage}[h]{1\linewidth}
			{\includegraphics[width=1\linewidth]{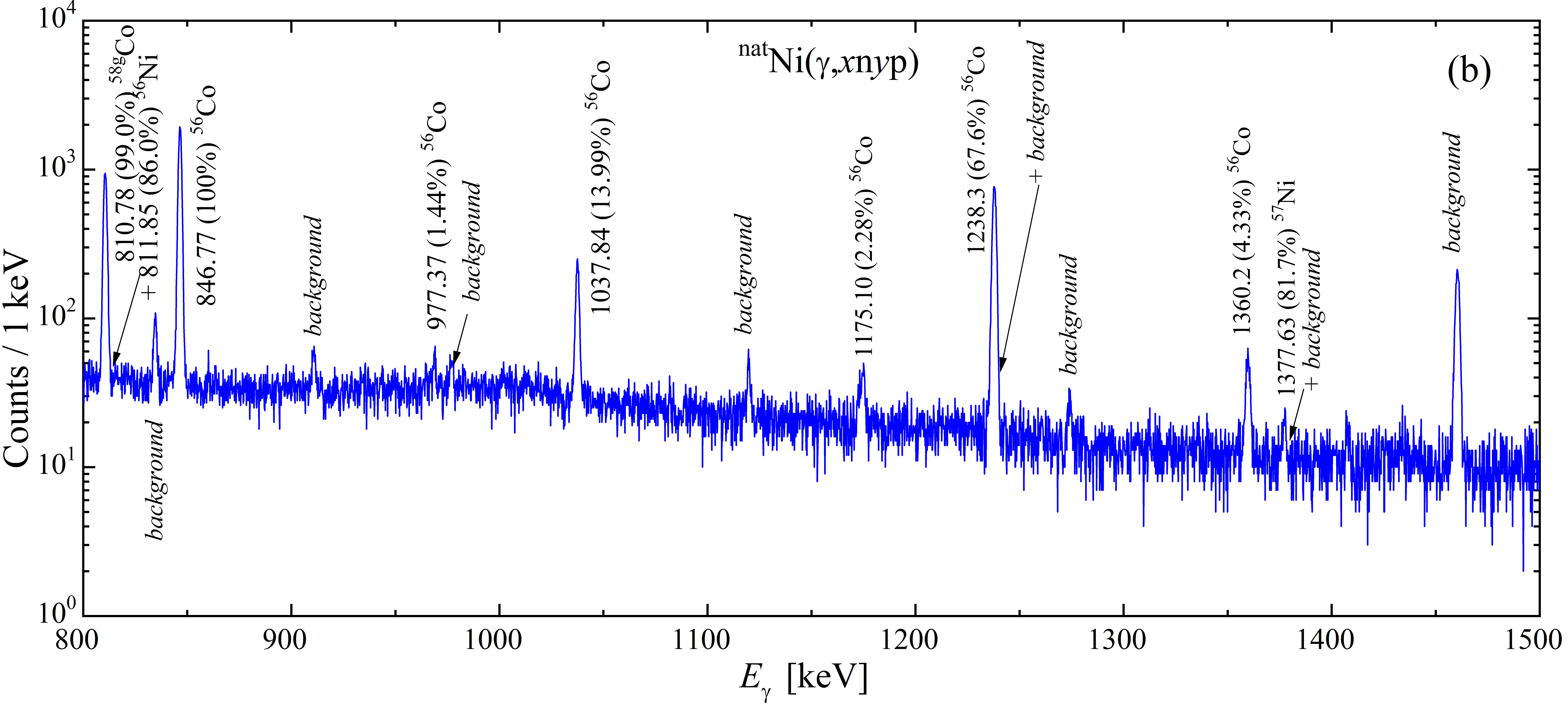}} \\
		\end{minipage}
		\caption{A spectrum of $\gamma$ rays emitted by the reaction products accumulated in the $^{\rm nat}$Ni target after bremsstrahlung $\gamma$-flux exposure with the energy $E_{\rm{\gamma max}}$ = 89.25~MeV. Relevant peaks are labeled with  the $\gamma$-ray transition energies in keV, and their intensities $I_{\gamma}$ (shown in parentheses). The  measurement time, $t_{\rm meas}$, and cooling time, $t_{\rm cool}$, were 84719 and 2925610 s, respectively.  }
		\label{fig3}
	\end{figure*}

	An energy spectrum of $\gamma$ rays emitted by the reaction products accumulated in the $^{\rm nat}$Ni target after bremsstrahlung gamma photon exposure is shown in Fig.~\ref{fig3}. The bremsstrahlung end-point energy was 89.25~MeV.
	
	In $^{\rm nat}$Ni targets, photoneutron reaction produce the $^{57}$Ni nuclei which decay via EC/$\beta^+$ process (100\%). We therefore observe cumulative activity of $^{57}$Co, which is the result of two processes: $^{\rm nat}$Ni($\gamma$,$x$np)$^{57}$Co and  $^{\rm nat}$Ni($\gamma$,$x$n)$^{57}$Ni $\xrightarrow{\rm{EC}/\beta^+}$ $^{57}$Co. The same applies to the production of the $^{55,56}$Co nuclei. Simplified decay schemes of the $^{55-57}$Co nuclei are shown in Figs. 4(a), 5(a), 6(a).  
	
	To determine the experimental flux-averaged cross-sections $\langle{\sigma(E_{\rm{\gamma max}})}\rangle$ 
	we used Eq.~(\ref{form2}) and the bremsstrahlung flux, integrated from the minimum reaction threshold $E^{\rm min}_{\rm thr}$ ($E_{\rm thr}$ in bold in Table~\ref{tab1}). Then, to take into account the difference between bremsstrahlung $\gamma$-flux from $E_{\rm{thr}}^{\rm min}$ and effective $\gamma$-flux, which causes the $^{55-57}$Co reaction products on all stable isotopes of $^{\rm nat}$Ni, these values were corrected with the coefficient defined as Eq. (1a)/Eq. (1b). 
	Corrected flux-averaged cross-sections are larger than results with $E^{\rm min}_{\rm thr}$ by 0.3--2.7~\% for investigated energies. 
	
	The obtained experimental cumulative flux-averaged cross-sections are listed in Tables B--D and graphically shown in Figs. 4(b), 5(b), 6(b).

	\subsubsection{ The photoproduction of $^{57}$Co on $^{\rm nat}$Ni }
	\label{sec:57Co} 
	
	To identify the $^{57}$Co nuclei, the 122.06 and 136.47~keV transitions with  their intensities of 85.60 and 10.68\%, respectively, were used.  As can be seen from Fig. 4(b), the results for two used transitions coincide within the experimental errors.

	The theoretical flux-averaged cross-sections $\langle{\sigma(E_{\rm{\gamma max}})}\rangle_{\rm th}$ were calculated using Eq.~(1a). In Fig. 4(b), $\langle{\sigma(E_{\rm{\gamma max}})}\rangle_{\rm th}$ for the $^{\rm nat}$Ni($\gamma$,$x$np)$^{57}$Co reaction are shown, as well as for cumulative process: $^{\rm nat}$Ni($\gamma$,$x$np)$^{57}$Co + $^{\rm nat}$Ni($\gamma$,$x$n)$^{57}$Ni $\xrightarrow{\rm{EC}/\beta^+}$ $^{57}$Co.
	As can be seen, the difference between the cumulative flux-averaged cross-section and cross-section for the $^{\rm nat}$Ni($\gamma$,$x$np)$^{57}$Co reaction is about 33\%.
	
	A comparison of the experimental data with the calculation performed for the cumulative flux-averaged cross-section for the production of  $^{57}$Co nucleus on $^{\rm nat}$Ni shows a good agreement (see Fig.~4(b)). 
	
	The calculated values of $Y_k(E_{\rm{\gamma max}})$ revealed that in the investigated energy region the $^{57}$Co nucleus is basically produced from the $^{58}$Ni isotope. At $E_{\rm{\gamma max}}$ = 94~MeV the $^{58}$Ni isotope contributed more than 98.3\% to the $^{58}$Ni($\gamma$,p) reaction. In the case of the production of the $^{57}$Ni nucleus on $^{\rm nat}$Ni, the main contribution (more than 99\%) also comes from the reaction on $^{58}$Ni, see \cite{55}.
	With the decrease of energy $E_{\rm{\gamma max}}$, these contributions smoothly increase and for lower energies $E_{\rm{\gamma max}}$ are close to 100\%.
	
	\begin{figure}[ht!]
		\begin{minipage}[h]{0.85\linewidth}
			{\includegraphics[width=1\linewidth]{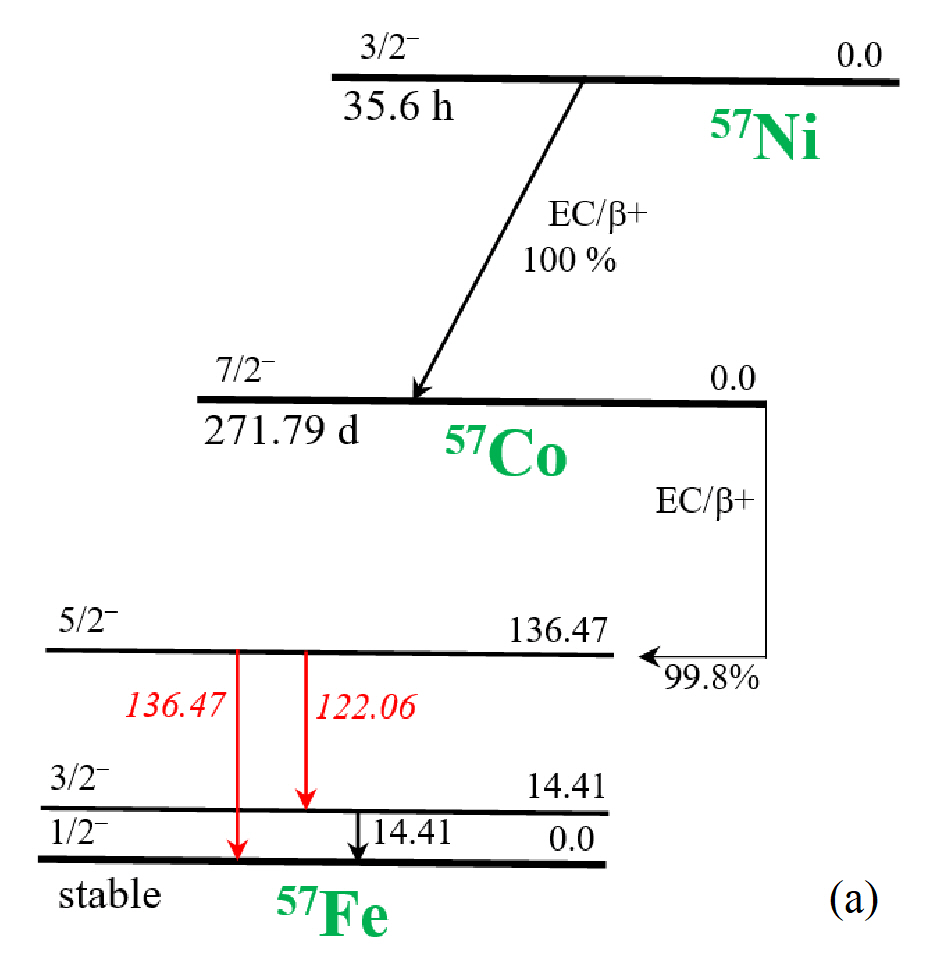}} \\
		\end{minipage}
		\vfill
		\begin{minipage}[h]{1.0\linewidth}
			{\includegraphics[width=1\linewidth]{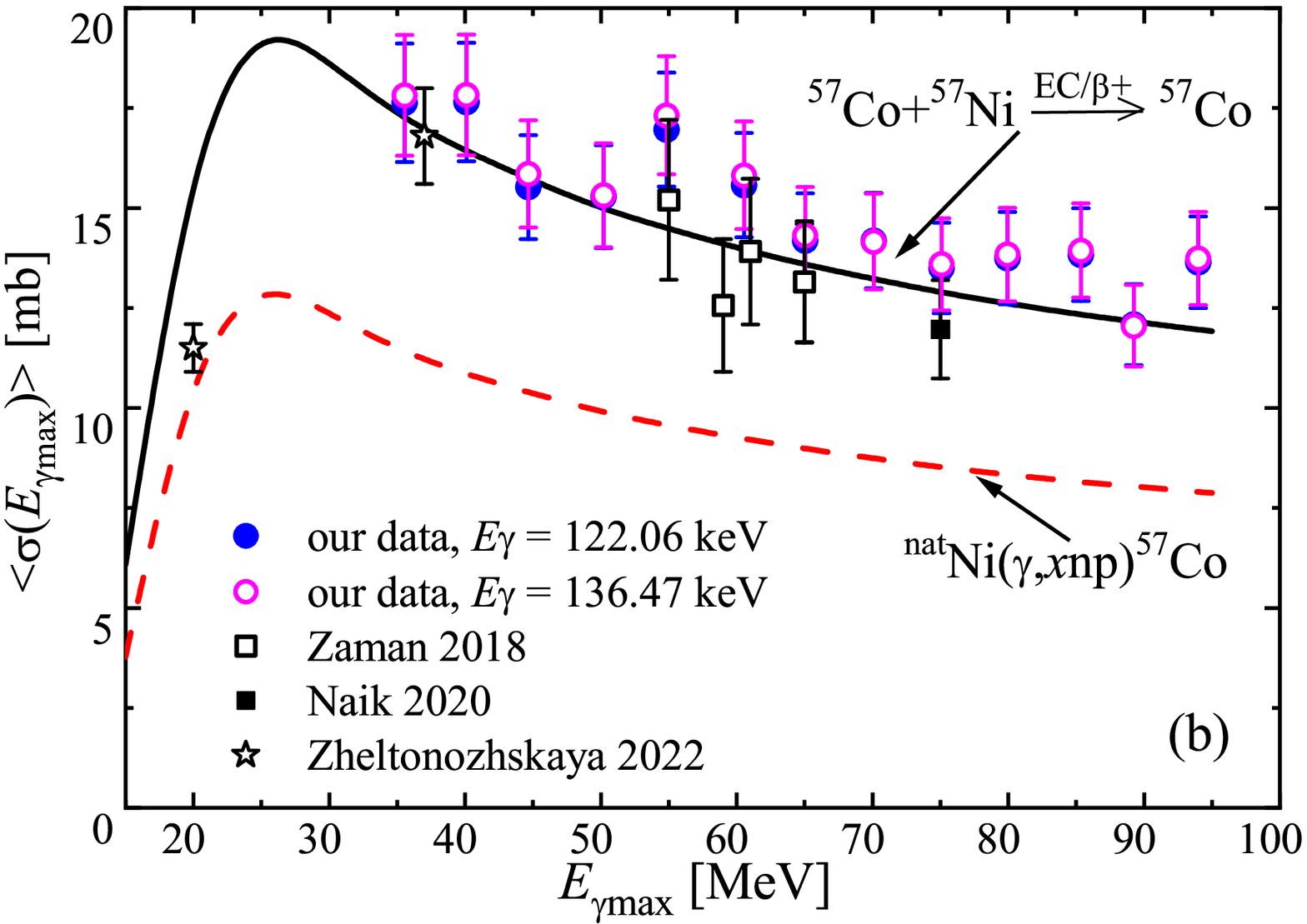}} \\
		\end{minipage}
		\caption{(a) Simplified decay scheme of $^{57}$Co. The nuclear level energies are in keV. The red-colour arrows represent the $\gamma$-ray transitions used in this work. 
			(b) The flux-averaged cross-sections $\langle{\sigma(E_{\rm{\gamma max}})}\rangle$ for the photonuclear production of $^{57}$Co on $^{\rm nat}$Ni. Experimental data: empty and full circles -- data from this work. 
			Literature values are adopted from Zaman et al. \cite{34}, Naik et al. \cite{35}, and Zheltonozhskaya et~al. \cite{Zhe}.
			Theoretical calculations, performed using  Eq.~(1a) and cross-sections from the  TALYS1.95 code: dashed curve -- $^{\rm nat}$Ni($\gamma$,$x$np)$^{57}$Co, solid curve -- cumulative cross-section.}
		\label{fig4}
	\end{figure}
	
	\begin{table}[h]
		\caption{\label{tab2} Experimental cumulative flux-averaged cross-section $\langle{\sigma(E_{\rm{\gamma max}})}\rangle$  [mb] of the photoproduction of $^{57}$Co on $^{\rm nat}$Ni and $^{58}$Ni. The results for the 122.06~keV transition are presented (see Fig.~(\ref{fig3}) and (\ref{fig4})). }
		\centering
		\begin{tabular}{ccc}
			\noalign{\smallskip}
			\hline	
			$E_{\rm{\gamma max}}$ (MeV) &  $^{\rm nat}$Ni($\gamma$,$x$np)$^{57}$Co  & $^{58}$Ni($\gamma$,p)$^{57}$Co* \\   \hline	\noalign{\smallskip}
			35.55 &  17.6~{\it  15} &  25.8~{\it  22}  \\ 
			40.10 &  17.7~{\it  15} &  25.5~{\it  21}  \\ 
			44.65 &  15.5~{\it  13} &  22.2~{\it  19}  \\ 
			50.20 &  15.3~{\it  13} &  21.7~{\it  18}  \\ 
			54.85 &  17.0~{\it  14} &  24.0~{\it  20}  \\ 
			60.55 &  15.6~{\it  13} &  22.0~{\it  19}  \\ 
			65.05 &  14.2~{\it  12} &  20.0~{\it  17}  \\ 
			70.10 &  14.2~{\it  12} &  20.0~{\it  17}  \\ 
			75.10 &  13.5~{\it  11} &  19.0~{\it  16}  \\ 
			79.95 &  13.7~{\it  12} &  19.4~{\it  16}  \\ 
			85.35 &  13.8~{\it  12} &  19.5~{\it  16}  \\ 
			89.25 &  12.1~{\it  10} &  17.0~{\it  14}  \\ 
			94.00 &  13.6~{\it  11} &  19.2~{\it  16}  \\ \hline   \noalign{\smallskip}			
		\end{tabular} \\ 
		\footnotesize{* For the calculation of the cumulative photoproduction of $^{57}$Co on $^{58}$Ni, the normalized reaction yield $Y_k(E_{\rm{\gamma max}})$ and the isotopic abundance of $^{58}$Ni isotope in $^{\rm nat}$Ni were used.}
	\end{table}

	\subsubsection{ The photoproduction of $^{56}$Co on $^{\rm nat}$Ni }
	\label{sec:56Co} 
	
	To identify $^{56}$Co nucleus, the 846.77, 1037.84 and 1238.28~keV transitions with  intensities of 99.9399, 14.05 and 66.46\%, respectively, were used.
	
	Since the cooling time was 32--53 days, not all of the $^{56}$Ni nuclei, formed in the target at the end of bombarding (EOB), decayed into $^{56}$Co. Our estimates showed that at $t_{\rm cool}$ = 32 days, 2.6\% of the $^{56}$Ni EOB nuclei remained in the target. With increasing cooling time, this value quickly decreases and at $t_{\rm cool}$ = 53 days it is only 0.24\%. The experimental data were not corrected for this contribution. However, to take into account the influence of the remaining $^{56}$Ni nuclei on cumulative yields, the experimental errors were increased accordingly.
	
	As known, the natural background spectrum contains $\gamma$ ray with $E_\gamma$ = 1238.48~keV. Since the induced activity spectra were recorded for 20--95 hours, the background radiation makes some additional contribution to the intensity of the $\gamma$-ray line with $E_\gamma$ = 1238.28~keV. Our estimates showed that the background contribution to the total peak value ranged from 2.2\% to 16.7\%. The larger contribution was at low energies $E_{\rm{\gamma max}}$ where the cross-section for the photonuclear reaction is smaller. The magnitude of the background contribution was estimated with an accuracy of no worse than 3.6\%, which is taken into account in the total error of the final results for $\langle{\sigma(E_{\rm{\gamma max}})}\rangle$ along the 1238.28~keV transition.
	
	The obtained experimental data of the bremsstrahlung flux-averaged cross-sections $\langle{\sigma(E_{\rm{\gamma max}})}\rangle$ for the photoproduction of $^{56}$Co are shown in Fig. 5(b). The results for all $\gamma$ rays coincide within the experimental errors. 
	
	All experimental data are slightly lower the calculation result for the cumulative flux-averaged cross-section (see  Fig.~5(b)). 
	As in the case of the $^{57}$Co nucleus, the photoproduction of $^{56}$Co nucleus in the energy region studied is basically produced from the $^{58}$Ni isotopes. The $^{58}$Ni isotope contributed more than 97.6\% to 
	$^{56}$Co production at $E_{\rm{\gamma max}}$ = 94~MeV. With the energy decreasing, this contridution smoothly increases and for lower energies $E_{\rm{\gamma max}}$ are close to 100\%. 
	
	\begin{figure}[ht!]
		\begin{minipage}[h]{0.85\linewidth}
			{\includegraphics[width=1\linewidth]{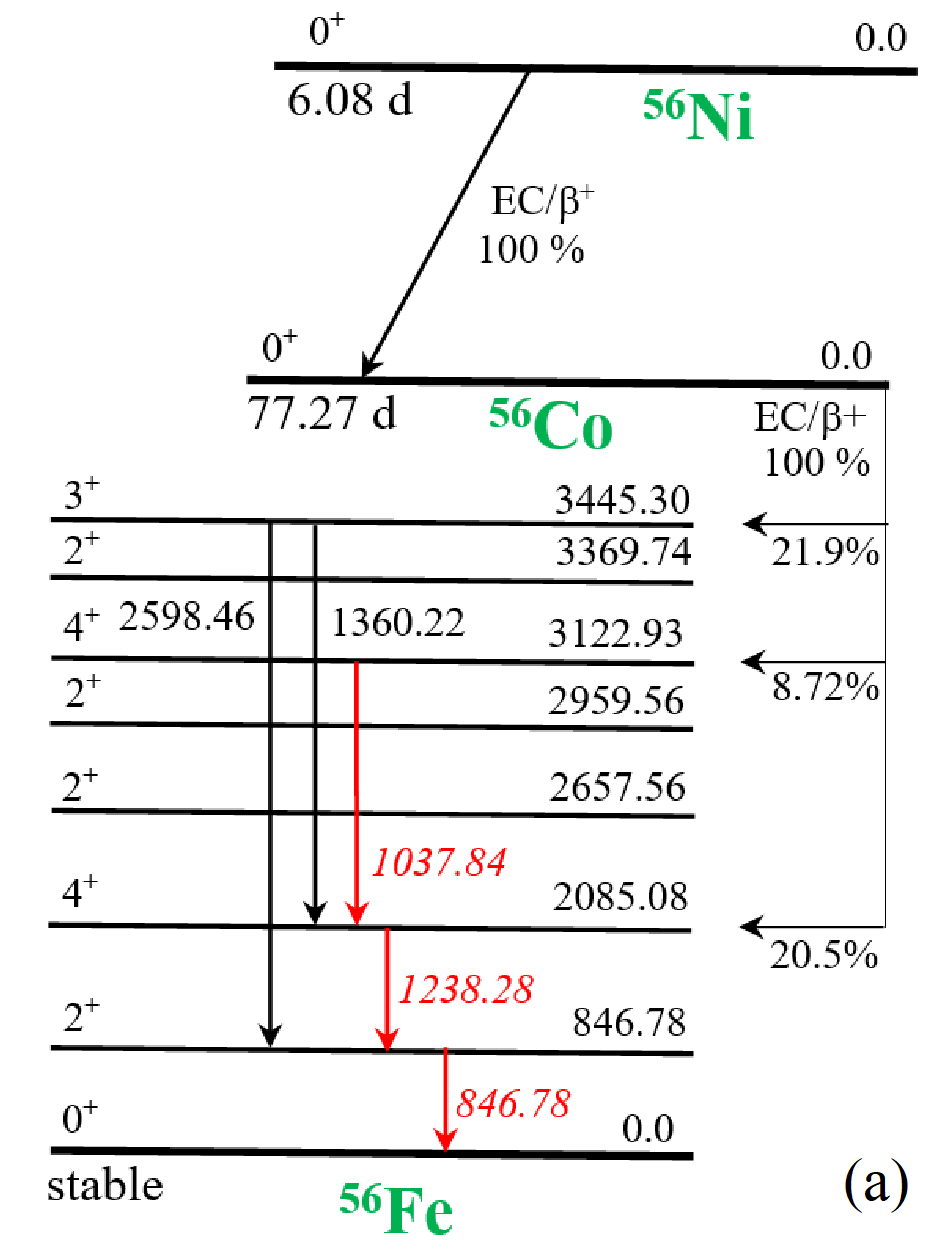}} \\
		\end{minipage}
		\vfill
		\begin{minipage}[h]{1.0\linewidth}
			{\includegraphics[width=1\linewidth]{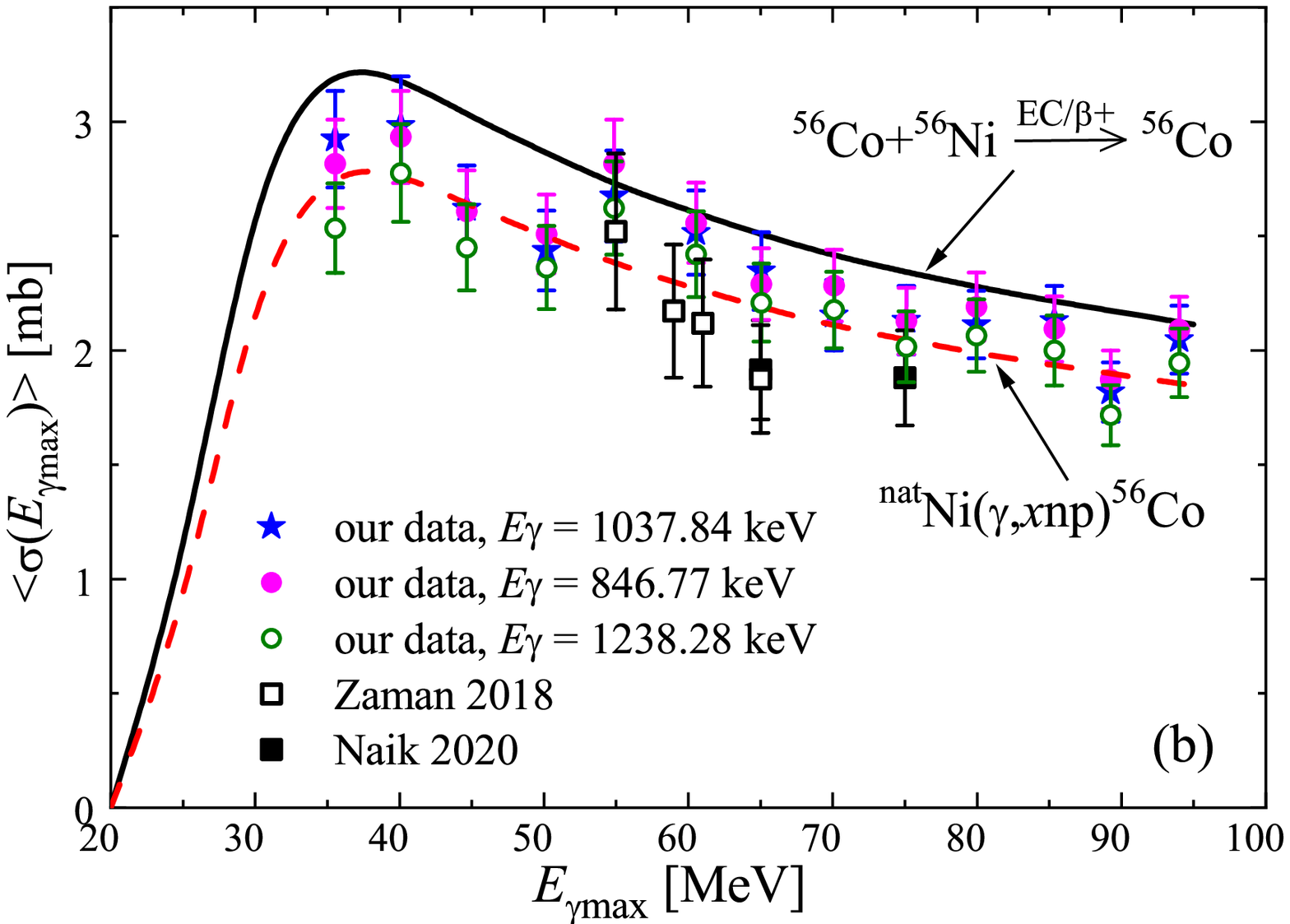}} \\
		\end{minipage}
		\caption{(a) Simplified decay scheme of  $^{56}$Co. 
			(b) The flux-averaged cross-sections $\langle{\sigma(E_{\rm{\gamma max}})}\rangle$ for the photonuclear production of $^{56}$Co on $^{\rm nat}$Ni. Experimental data: circles and stars -- data from this work. Literature values are adopted from Zaman et al. \cite{34} and Naik et al. \cite{35}. Curves are theoretical calculations. }
		\label{fig5}
	\end{figure}
	
	\begin{table}[h]
		\caption{\label{tab3} Experimental cumulative flux-averaged  cross-section $\langle{\sigma(E_{\rm{\gamma max}})}\rangle$  [mb] of the photoproduction of $^{56}$Co on $^{\rm nat}$Ni and $^{58}$Ni. The results for the 846.77~keV transition are presented (Fig.~(\ref{fig3}) and (\ref{fig5})).   }
		\centering
		\begin{tabular}{ccc}
			\hline 
			$E_{\rm{\gamma max}}$ (MeV) &  $^{\rm nat}$Ni($\gamma$,$x$np)$^{56}$Co  & $^{58}$Ni($\gamma$,pn)$^{56}$Co*  \\   \hline	\noalign{\smallskip}
			35.55 & 2.82~{\it  19} &  4.14~{\it  28}  \\ 
			40.10 & 2.93~{\it  20} &  4.31~{\it  30}   \\ 
			44.65 & 2.61~{\it  18} &  3.83~{\it  26}  \\ 
			50.20 & 2.51~{\it  17} &  3.67~{\it  25}   \\ 
			54.85 & 2.82~{\it  19} &  4.07~{\it  28}   \\ 
			60.55 & 2.56~{\it  18} &  3.66~{\it  25}  \\ 
			65.05 & 2.29~{\it  16} &  3.25~{\it  22}  \\ 
			70.10 & 2.28~{\it  16} &  3.23~{\it  22}  \\ 
			75.10 & 2.13~{\it  15} &  3.00~{\it  20}  \\ 
			79.95 & 2.19~{\it  15} &  3.08~{\it  21}   \\ 
			85.35 & 2.09~{\it  14} &  2.93~{\it  20}    \\ 
			89.25 & 1.87~{\it  13} &  2.62~{\it  18}    \\ 
			94.00 & 2.09~{\it  14} &  2.92~{\it  20}    \\ 	 \hline \noalign{\smallskip}		\end{tabular} \\ 
		\footnotesize{* as in the case of the $^{58}$Ni($\gamma$,p)$^{57}$Co reaction, see Table \ref{tab2}.}
	\end{table}
	
	\subsubsection{ The photoproduction of $^{55}$Co on $^{\rm nat}$Ni }
	\label{sec:55Co} 
	
	To identify $^{55}$Co it was used the  $\gamma$-ray transition with $E_\gamma$ = 931.1~keV and the intensity of 75.0\%.  
	
	The theoretical flux-averaged cross-sections $\langle{\sigma(E_{\rm{\gamma max}})}\rangle_{\rm th}$ for the $^{\rm nat}$Ni($\gamma$,$x$np)$^{55}$Co reaction,  as well as for cumulative process are shown in Fig. 6(b).
	As can be seen, the difference between these theoretical estimates,  direct and cumulative, is less than 1\%.
		The obtained experimental cumulative cross-sections $\langle{\sigma(E_{\rm{\gamma max}})}\rangle$  are significantly lower than the theoretical estimates. 
	
	The $^{55}$Co nucleus in the investigated energy region is basically produced from the $^{58}$Ni isotopes. 
	The $^{58}$Ni isotope contributed more than 98\% to $^{55}$Co photoproduction at $E_{\rm{\gamma max}}$ = 94~MeV.
	With the energy decreasing, this contribution smoothly increases and for lower energies $E_{\rm{\gamma max}}$ are close to 100\%. 
	The reaction $^{58}$Ni($\gamma$,2np)$^{55}$Co accounts for more than 98\% of the total production of the $^{55}$Co nucleus. Thus, the observed difference between the experimental result and theoretical prediction is related to the accuracy of the calculation of this reaction. 
	
	These are experimental results for photoproduction of the $^{55-57}$Co isotopes on $^{\rm nat}$Ni of other authors \cite{34,35,Zhe}. These data were obtained using the $\gamma$-activation method.
	In all three cases the obtained experimental cross-sections are in good agreement with those available in the literature (see Figs.~ 4(b), 5(b) and 6(b)).
	
	\begin{figure}[ht!]
		\begin{minipage}[h]{0.85\linewidth}
			{\includegraphics[width=1\linewidth]{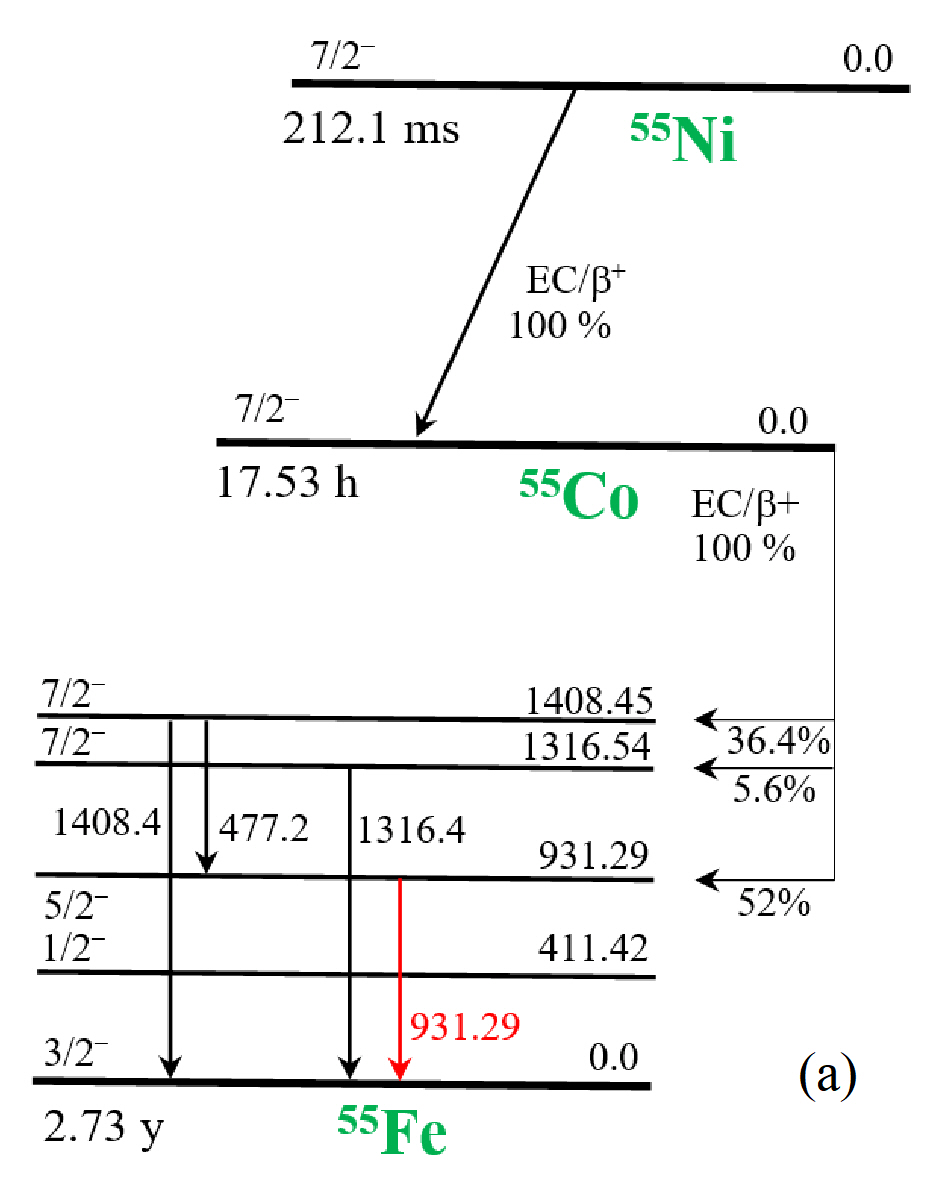}} \\
		\end{minipage}
		\vfill
		\begin{minipage}[h]{1.0\linewidth}
			{\includegraphics[width=1\linewidth]{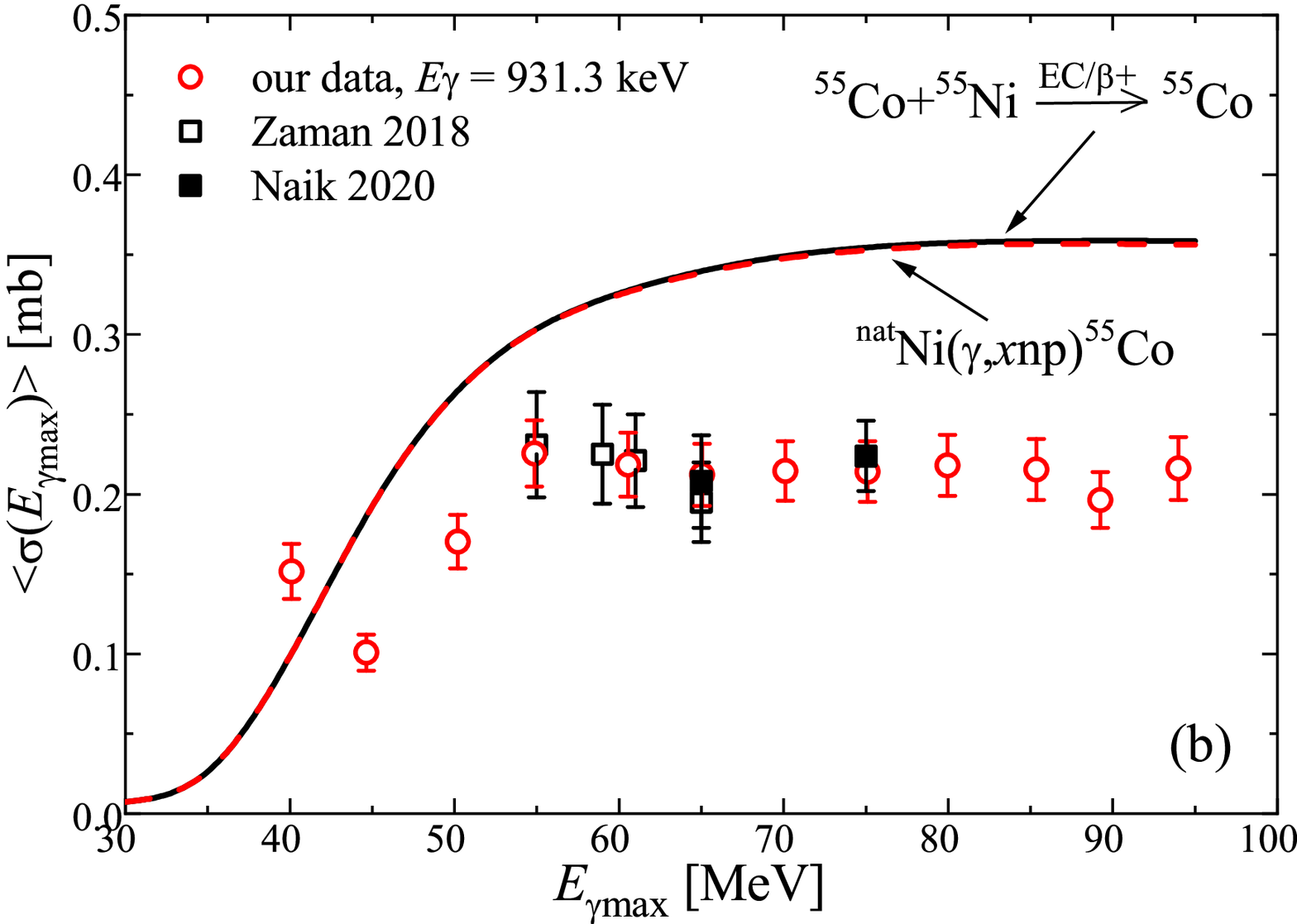}} \\
		\end{minipage}
		\caption{(a) Simplified decay scheme of $^{55}$Co. 
			(b) The flux-averaged cross-sections $\langle{\sigma(E_{\rm{\gamma max}})}\rangle$  for the photonuclear production of $^{55}$Co on $^{\rm nat}$Ni.  Experimental data: circles -- data from this work. Literature values are adopted from Zaman et al. \cite{34} and Naik et al. \cite{35}. Curves are theoretical calculations.}
		\label{fig6}
	\end{figure}
	
	\begin{table}[h]
		\caption{\label{tab4} Experimental cumulative flux-averaged cross-section $\langle{\sigma(E_{\rm{\gamma max}})}\rangle$  [mb] of the photoproduction of $^{55}$Co on $^{\rm nat}$Ni and $^{58}$Ni. The results for the  931.3~keV transition are presented (Figs.~(\ref{fig3}) and (\ref{fig6})). }
		\centering
		\begin{tabular}{ccc}
			\noalign{\smallskip} \hline
			$E_{\rm{\gamma max}}$ (MeV) &  $^{\rm nat}$Ni($\gamma$,$x$np)$^{55}$Co  & $^{58}$Ni($\gamma$,2np)$^{55}$Co* \\   \hline	\noalign{\smallskip}
			40.10 & 0.152~{\it  19} &  0.223~{\it  28}  \\ 
			44.65 & 0.101~{\it  12} &  0.148~{\it  18}  \\ 
			50.20 & 0.170~{\it  19} &  0.250~{\it  27}  \\ 
			54.85 & 0.226~{\it  23} &  0.331~{\it  34}  \\ 
			60.55 & 0.219~{\it  22} &  0.320~{\it  32}   \\ 
			65.05 & 0.212~{\it  22} &  0.308~{\it  31}   \\ 
			70.10 & 0.215~{\it  21} &  0.309~{\it  31}   \\ 
			75.10 & 0.214~{\it  22} &  0.306~{\it  31}   \\ 
			79.95 & 0.218~{\it  22} &  0.310~{\it  31}    \\ 
			85.35 & 0.215~{\it  22} &  0.305~{\it  31}     \\ 
			89.25 & 0.196~{\it  20} &  0.277~{\it  28}     \\ 
			94.00 & 0.216~{\it  22} &  0.305~{\it  32}     \\ \noalign{\smallskip}	\hline		\end{tabular} \\ 
		\footnotesize{ * as in the case of the $^{58}$Ni($\gamma$,p)$^{57}$Co reaction, see Table \ref{tab2}.}
	\end{table}

	\section{Conclusions}
	\label{Concl}
	
	In this work, the flux-averaged cross-sections $\langle{\sigma(E_{\rm{\gamma max}})}\rangle$ of the production of $^{55-57}$Co in photonuclear reactions on $^{\rm nat}$Ni have been measured in the bremsstrahlung end-point energy range $E_{\rm{\gamma max}}$ = 35--94~MeV. The obtained experimental results  were compared with the data available in the literature, and were in a good agreement. 
	
	The theoretical flux-averaged cross-sections $\langle{\sigma(E_{\rm{\gamma max}})}\rangle_{\rm th}$ were estimated using the cross-section values $\sigma(E)$ from the TALYS1.95 code. The cumulative experimental and theoretical cross-sections satisfactory agree in the case of the production of  $^{57}$Co and  $^{56}$Co. For the reactions with the production of  $^{55}$Co nuclei, the theoretical values differ from experimental ones. It has been shown that the main contribution (more than 97\%) to the production of  $^{55-57}$Co nuclei in photoinduced reactions on natural nickel comes from $^{58}$Ni.
	
	The obtained experimental results expand the energy range of previously known data for the cumulative results of $^{\rm nat}$Ni($\gamma$,$x$np)$^{55-57}$Co and can be used to test various theoretical models.

\section*{Acknowlegment}
	The authors would like to thank the staff of the linear electron accelerator LUE-40 NSC KIPT, Kharkiv, Ukraine, for their cooperation in the realization of the experiment.
	
	This work was supported by the Slovak Research and Development Agency under No. APVV-20-0532 and APVV-22-0304, and the Slovak grant agency VEGA (Contract No. 2/0175/24). Funded by the EU NextGenerationEU through the Recovery and Resilience Plan for Slovakia under the project No. 09I03-03-V01-00069.

\end{document}